# Composing vibrotactile music: A multi-sensory experience with the Emoti-chair

Anant Baijal, Julia Kim, Carmen Branje, Frank Russo, Deborah I. Fels

# Composing vibrotactile music:
# A multi-sensory experience with the Emoti-Chair

Anant Baijal[Ω]  Julia Kim[#]  Carmen Branje[*]  Frank Russo[*]  Deborah I. Fels[*]

VIT University, India

McMaster University, Canada

Ryerson University, Canada

Ryerson University, Canada

Ryerson University, Canada

**ABSTRACT**

The Emoti-Chair is a novel technology to enhance entertainment through vibrotactile stimulation. We assessed the experience of this technology in two workshops. In the first workshop, deaf film-makers experimented with creating vibetracks for a movie clip using a professional movie editing software. In the second workshop, trained opera singers sang and 'felt' their voice through the Emoti-Chair. Participants in both workshops generally found the overall experience to be exciting and they were motivated to use the Chair for upcoming projects.

**INDEX TERMS:** H.1.2 [User/Machine Systems]: Human Factors; H.5.2 [User Interfaces]: Haptic I/O; H.5.5 [Sound and Music Computing]: Methodologies and techniques; I.2.10 [Vision and Scene Understanding]: Video analysis; H.5.m [Information Interfaces and Presentation]: Miscellaneous- Tactile music; K.4.2 [Social Issues]: Assistive technologies for persons with disabilities

## 1   INTRODUCTION

Music has been observed in virtually all cultures ever studied and is thus arguably one of the most universal forms of entertainment. It is intertwined with our society and knits together various art forms. Music has many purposes for individuals including emotional health/well-being, aesthetics, social/cultural participation and development, therapy, relaxation and pure enjoyment [1]. This level of implication of music in most people's lives also means that it can be potentially used to manipulate or influence individuals on those levels. A good example is how merchandisers use music to influence buying behavior [2]. A second example is how music is used in film and television to support the flow of the narrative [3], set the mood and manipulate the audience through emotion [4]. Today, movie making is a multimodal form of entertainment relying upon audio and visual stimuli and can be considered incomplete without the element of music in it.

As music has been identified as having such a critical and influential role in society, in general, and in entertainment media and culture, in particular, one could assume that access for all citizens is important. However, individuals in the hard of hearing and the deaf communities (HH/D) are at a considerable disadvantage and have been virtually excluded in discussion of the uses and importance of music in society. While some members of these communities are not interested in having access to a hearing artifact such as music, many want to be included as a matter of equity and participation. Some technologies have been proposed to address the issues of music accessibility but most are either incapable of the same level of expression (e.g. closed captioning [5], have limited audiences (e.g., sign language [6]), or are confined to research laboratories (e.g., animated text [7]). While sign language may have the performance and lyrical expressiveness to represent music [8], it has a limited audience so that people who are hard of hearing would likely be excluded if only sign language was used to provide an alternative representation to music.

In addition, all of these technologies make use of the visual system. Visual representation of music as words in the case of captioning or animation in the case of animated text or graphics has the added complication of modality translation or substitution. It also requires cognitive interpretation of the meaning of the words which adds to the information processing workload. A modality substitution between two more closely aligned modalities such as the tactile and auditory systems may be less effortful and more efficient.

One less explored possibility for providing alternative access to music is the potential of the tactile system. The physiology and processing systems for hearing and tactile perception have many more similarities than the vision and hearing system. It may be possible to take advantage of the parallels for conveying music. [9] used a haptic chair and music visualization techniques in an attempt to convey music to deaf or hard of hearing people. In this study a single music signal was conveyed through music speakers placed on the feet, hands and back of a deaf person. One of the important findings of this research was the intensity of the response by the study participants to the haptic system. This may indicate that not only was there a strong desire to have access to music but that the tactile system may be the preferred modality with which to experience those desirable and emotional elements of the music. It may be a stronger experience of the aesthetic aspects of music rather than the intellectual exercise of reading text descriptions or understanding random graphic representations as is the case for captioning or music visualizations.

---

[Ω] School of Electrical Engineering, VIT University, Vellore, Tamil Nadu, 632 014, India (anantbaijal2008@vit.ac.in)

[#] McMaster University, Hamilton, Ontario, L8S 4L8, Canada (julia.sh.kim@gmail.com)

[*] Ted Rogers School of Information Technology Management, Ryerson University, 350 Victoria Street, Toronto, Ontario, M5B 2K3, Canada (cbranje@ryerson.ca, russo@psych.ryerson.ca, dfels@ryerson.ca )

A second system that has been recently developed to use the tactile system to express music is The Emoti-Chair [10]. It also allows people 'feel' the music through vibrations produced by an array of voice coils. It separates music into different frequency bands and then applies these frequencies on different areas of the human back using a chair form factors. Audience reactions (hearing and deaf) to the Emoti-Chair have been mostly positive [10], similar to the responses in [9]. One unanticipated outcome from the user studies is the interest shown by artist communities in exploring how to use this new technology to create new music forms of art as well make their sound art more accessible to people who are deaf or hard of hearing.

In this paper we report on a study engaging, professional film-makers and singers in creating and experiencing tactile music on the Emoti-Chair. This study explores compositional and performance strategies that artists take in creating what we call "vibe-tracks" or tactile music that audiences can 'feel' on The Emoti-Chair without sound.

We report on responses to pre and post questionnaires that collected participant views about the workshop and about vibrotactile stimulation in general. The workshop events were video recorded and then thematic categorization was performed on the clips. The workshops helped the research team in gaining feedback about the use of The Emoti-Chair and in carrying qualitative evaluation based on the participants' involvement.

## 2 BACKGROUND

Studies have shown that there is more to music than just 'hearing' and that music is a multisensory experience [11] [12]. It is also well known that during concerts, hard of hearing or deaf people often stand near loud speakers to infer sounds through vibrations, known as speaker listening, and that deaf or hard of hearing people often experience music as vibration rather than as sound [13]. There is a recognized relationship between the auditory and tactile perceptual systems and the manner in which they influence each other [14] [15]. As a result there is an opportunity to expand the use of the tactile system to represent more detailed sound information than what can be currently offered through stereo speakers and mixed sound tracks.

### 2.1 Sensory substitution and crossmodal interfaces

Crossmodal interfaces present information intended for one type of display using a different form of display, such as using vibrations to represent sound [16], or sound to represent images [17]. Multimodal interfaces incorporate multiple forms of input and output to provide a variety of devices to support human-computer interactions. Much of the extensive literature on multimodal and crossmodal interfaces can be divided into three main categories, multimodal input (e.g. eye gaze, gesture, speech, and standard computer input devices [18], [19]), output (visual, auditory, tactile [20]) and dialogue frameworks and architectures for integrating and mapping i/o information and signals with user commands, needs, or tasks [21]. In addition, there is much literature on design guidelines and recommendations for producing multi or crossmodal systems (see [22] for an overview). Much of this research is focused on supporting work tasks that are visually oriented, towards efficient task completion through information redundancy or substitution, rather than on creating psychological and physiological mappings that can replace information displayed through one modality *completely* using another [22].

While the technical aspect of crossmodal interface design presents interesting technical and theoretical challenges, thinking about how to represent musical concepts using the tactile modality, tactile music, represents an artistic or creative challenge that few have considered. Gunther [23] is one of the few artists who has considered tactile music and the complexity of expressing musical concepts with a lower fidelity sensory system (the auditory system has about 40 auditory filters spaced in 1/3 octave bands compared with only 4 channels for the vibrotactile system [24]). He also describes a few other art oriented projects that had a tactile component to them but most of these project involved substitution or augmentation of visual displays with haptic or tactile feedback.

### 2.2 Music composition

There have been numerous studies that have attempted to examine and model the process of music composition as either a set of linear processes that involve preparation, incubation, illumination and verification of the creative activity (adaptation of stage theory proposed by [25]) or a problem solving activity [26] that involves iterations of breaking the problem into goals, sub-goals and pathways and then attempting to find solutions. This occurs in reflexive and iterative processes [26]. In a longitudinal case study over a three year period, [27] found that composition is a process of recursive problem solving (information processing theory) along with high level monitoring and structure, and influenced or changed by occasions of creative insight or "ah ha" moments (Gestalt theory) where new directions would be taken.

Novice and expert composers have been found to use this type of compositional process but with different cycle periods and grain sizes. [28] Found that novices tended to focus on note-by-note structure whereas experts tended to organize activities and tasks at a higher level where information is grouped into larger structures (e.g., phrases) and overall goals drive the compositional process. In our study, we used expert composers/artists but they were asked to create music for the tactile system, a novel concept for this group of artists. The tactile modality made available to these artists contained unique compositional elements such as spatial arrangement and controllable vibrations as well as new limitations including a smaller range of possible "notes" compared with a musical/electronic "instrument." Further, we suggest that tactile music requires a new way of thinking about abstract musical properties such as rhythm and melody and how to express those properties. In this case, a composer could then be considered an expert in musical creativity but a novice in the expression of that creativity using the tactile modality. We designed a set of supporting mechanisms including the output system, called The Emoti-Chair, to assist artists in conceptualizing and creating tactile music.

### 2.3 The Emoti-Chair

The design of the chair is based on the Model Human Cochlea (MHC) [12] which uses vibration as sensory substitution for auditory signals and perception. Music input to the chair is distributed to the user via a set of 16 voice coils which are mounted on the seat and back of a chair form factor. If the music signal is a mixed signal such as a .mp3 file, the signal is frequency segmented into eight unique frequency bands representing frequencies between 27.5Hz and 1,000Hz (refer to [12] for an explanation of the translation algorithm). The low and high frequencies are mapped on the bottom and top portion of the chair respectively in adherence to the psychological literature that suggests that human humans tend to represent higher sounds with higher spatial positions relative to lower sounds [29] [30]. MAX/MSP is used for the chair-computer interface to control power levels (volume) to each voice coil and to enable the entire system including the assignment of frequency bands to various voice coils.

The Emoti-Chair (see Figure 2) was earlier demonstrated in many different public venues such as live concerts, and as an installation at a well-known science centre. These venues allowed people who were deaf or hard of hearing to experience sound-based music using the tactile domain (see [10] for reports and analysis of the user experience of these events). However, the workshops described in this paper were the first time that professional singers and film-makers produced vibe-tracks (tactile music) for the HH/D in a real time setting. The purpose of these workshops was to determine whether the Emoti-Chair could create vibrotactile music and also to gain insight into professional artists' experience while creating such music. The reader is referred to [10] for detailed description of The Emoti-Chair.

## 3 METHODOLOGY

### 3.1 Research Questions

We report on an exploratory study of two workshops, one in Vancouver and one in Toronto, on tactile music creation (the workshop protocol was approved by the Ryerson Ethics board). The purpose of these workshops was to examine the processes that would be involved in creating tactile music. Given that this activity involves a new sensory element that most artists have never conceived of, we wanted to determine how groups of artists might approach the design/creative space with few constraints. Further we wanted to determine what impact musical expertise may have on that process. Towards this very ambitious goal, the following research questions were considered:

1) What are the reactions of artists to the concept and possibility of vibrotactile music?
2) What are the creative possibilities of vibrotactile music?

### 3.2 Participants

The participants from the Classical Singer workshop held in Vancouver consisted of hearing, professional vocalists and music professors (two male and ten female). The participants from the Film-maker workshop held in Toronto were deaf, professional film-makers (two male and four female) from various cities across North America. Most of the participants had between fifteen and twenty years of producing and editing films or performing music.

### 3.3 Classical singer workshop setup

To begin the workshop the Emoti-Chair was introduced to all participants by providing a brief overview of the theory and operation of the design (note: all participants were unpaid). The next component of the workshop consisted of allowing participants to explore the performance and compositional space afforded by the tactile modality and the Emoti-Chair. Two Emoti-Chairs were then setup with microphones so that participants could sing into the microphones and feel their own voices as well as the person's voice in the second chair. Following this brief introduction to the concept of tactile music, participants were divided into three groups of three to four people each and asked to have one person sing and the others feel the vocalisations in the chair. They were asked to sing familiar songs as well as experiment with original material.. Three Emoti-Chairs were set-up with two microphones (attached to microphone stands) per each chair; however one of the chairs experienced a technical problem and was removed from use in the workshop. In order to mute out any external sounds, which may bias their experience with the physical vibrations of the Emoti-Chair, the participant sitting on the chair could use headphones that produced loud white noise. However, the singers would have still felt vibrations through their facial bones, vocal cords and body.

Following their experimentation, which lasted approximately one hour, participants were asked to provide comments on their experience and then complete a post-study questionnaire. The post-study questionnaire consisted of 20 questions (six open ended, 14 forced choice/Likert scale questions). Two questions related to the workshop itself, one question asked about the enjoyment of feeling the vibrations, five questions asked specifically about using or experiencing the Emoti-Chair and the remaining questions asked about the process of composing vibrotactile music (learning how, becoming proficient at composing/performing vibrotactile music, ease of composing/performing, genres of music best suited to vibrotactile music, and predicting time for learning and becoming proficient at vibrotactile music for others).

The entire workshop was video recorded for later analysis.

### 3.4 Film-maker workshop setup

During the Film-maker workshop, six Deaf film-makers, each paid $500 for their participation each took part in the half day workshop. Participants were divided into two groups of three. Each group had one Emoti-Chair and a desktop computer installed with video editing software. An Adobe Premiere CS4™ project with a single video track and eight audio tracks, each of which was associated to the each of the vibrotactile channels of The Emoti-Chair was provided to participants. To support the compositional process, a set of vibrotactile segments that could be considered analogous to a "melodious phrase" and similar to the idea of 'tune blocks' used by [31] were developed. The tune blocks consisted of twenty vibrotactile segments, eighteen of which were comprised of sine or square waves of varying frequencies and two were frequencies sweeps. The frequencies included for the sine waves were 40 Hz, 50 Hz, 60 Hz, 70 Hz, 80 Hz, 90 Hz, 100 Hz, 105Hz, 110 Hz, 125 Hz, 150 Hz, 175 Hz, 200 Hz, 250 Hz, 300 Hz and 350 Hz. A square wave and a saw tooth wave were provided at 60 Hz. Two frequencies sweeps were also provided, one nine seconds in length, sweeping from 40 Hz to 350 Hz and the other five second in length and sweeping from 100 Hz to 300 Hz. The vibrotactile segments could be placed along the time line in Premiere project, and where the vibration is presented could be determined by placing the segments on specific tracks (see Figure 1).

The group task was to create vibetracks for a pre-selected two-minute movie clip from the movie, Harry Potter and the Order of Phoenix™ (2007) using Adobe Premiere CS4. This scene showed the protagonist and supporting characters fiercely battling against numerous antagonists. The scene contained obvious mood elements of suspense, tension and action. After the experimentation sessions, participants from both groups were engaged in post-experimentation discussions. The entire workshop was video recorded for later analysis.

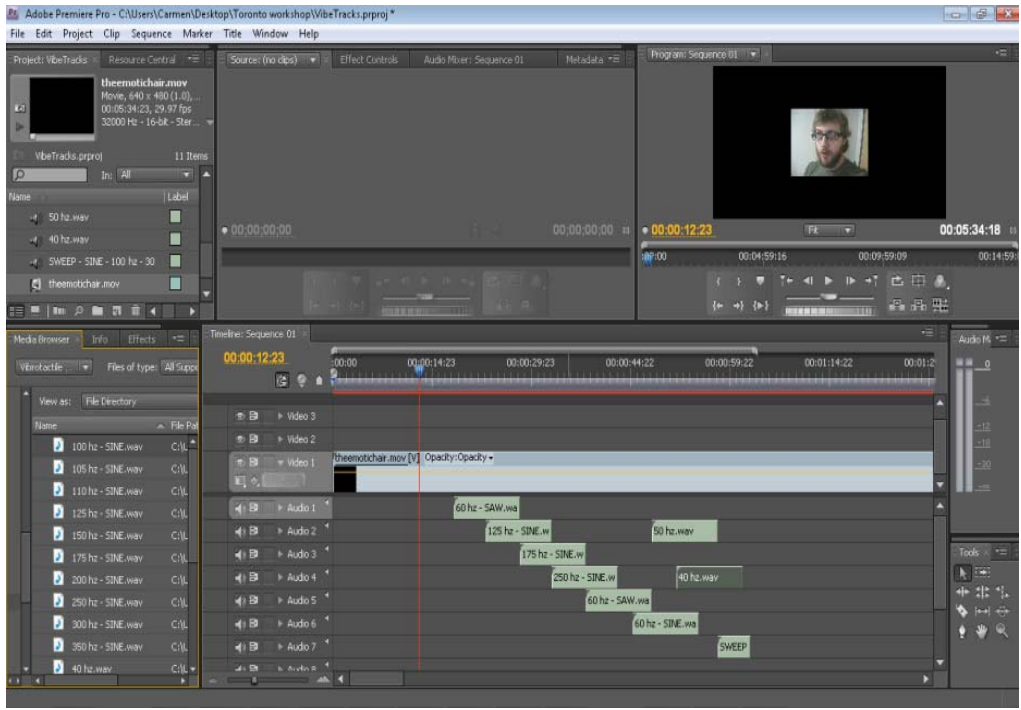

Figure 1: An example of how a Premiere Pro project was set up to facilitate the creation of vibetracks for a film. A single video track is present along with eight audio tracks, each associated to an area of The Emoti-Chair. The vibrational segments can be seen placed on the multiple audio tracks.

## 3.5 Data analysis

The video recordings of both workshops were analyzed to examine how professional artists use and explore vibetracks and how it might be used in their particular artist domain.

We used professional, event logging software, Observer XT 9.0, to analyse the video recordings of the Classical Singer workshop (1:24:23 hrs) and the Film-maker workshop (2:28:09 hrs). The portion of the video analysed in the paper represents the part of the workshop where participants actually used the Emoti-Chair and made qualitative comments.

We carried out a thematic analysis as outlined by [32] and identified a set of six themes post hoc (see Table 1 for definitions). Two independent researchers then coded a randomly selected ten-minute clip of each workshop. An Interclass Correlation Coefficient (ICC) value was found to be 0.72 indicating that there was a sufficient correlation between the two raters. A single rater then continued coding membership in categories and duration of comments for each video.

### Thematic Classification Categories

**Technical comments** - Statements not directly related to the performance of the Emoti-Chair, such as any issues and problems with: room, environment (temperature and light conditions), chair comfort, microphones, computer, crowd, and workshop logistics (length and location).

**Emotional responses** - Words or statements highlighting the emotions experienced while using the Emoti-Chair. Negative emotions include feelings of frustration, sadness, anger and any words describing disinterest. Positive emotions involve feelings of excitement, and happiness, with words describing interest and likable qualities.

**Comments on vibrational aspects of the chair** - Statements regarding the physical feeling of the chair, specifically through the vibrations. Comments are both positive and negative. Vibrations can be described in terms of intensity/amplitude and frequency.

**Expressions of improvement/wishes** - Statements concerning the Emoti-Chair and its related experience or chair-specific components of software. The suggestions are in regards to the Emoti-Chair, vibetrack, and alternative implementation of the vibrational technology. May express as "I wish…" or "I could help…" and express the desire for more workshops to continue working with the chair.

**Shift in thinking** - Statements expressing both positive and negative comments regarding the newness, creativity and innovation of the Emoti-Chair. Positive statements may include expressions like: "This is amazing", "Changes how I think about my work", "Now I understand". Negative statements may include expression of non-innovation, pessimism, and the level of difficulty in conceptualizing and in learning the technical aspects of using the chair.

**Accessibility comments** - Statements expressing the use of Emoti-Chair for different audience apart from the workshop, such as hearing/hard-of-hearing, autistic, deaf, different age group, patients and other professionals from different job fields, such as actors, musicians, psychologists, and etc. The statements also express suggestion in use of the chair in different context, such as in theatre, educational setting, and healthcare.

Table1. Definitions for thematic classification

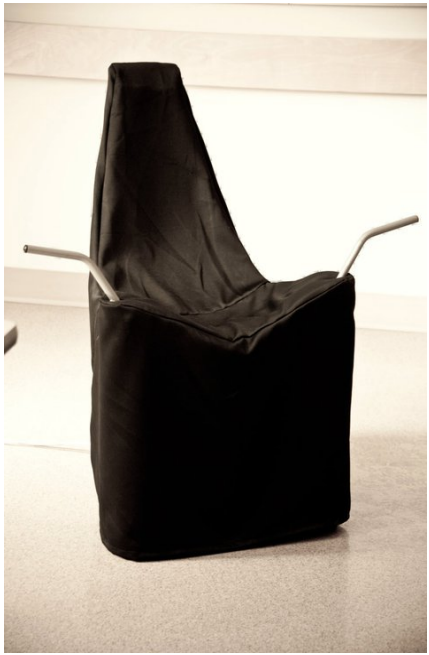

Figure 2: The Emoti-Chair

## 4 RESULTS AND DISCUSSION

### 4.1 Post-study questionnaire

A Mann-Whitney non-parametric test was used to compare responses to the post-study question between the Film-maker and Classical singer participants. There was no significant difference observed in any of the questions. In general, both groups seemed to have similar responses; however, more subjects were required to fully examine any statistical differences in responses to a post-study questionnaire.

Descriptive results from the questionnaire provide some insight into the similarities and possible differences. A Likert scale was used for the questions about the experience with the Emoti-Chair where 1 represents 'strongly disagree' and 5 represents 'strongly agree'. The film-makers and vocalists/music composers agreed that they enjoyed feeling the vibrations in the Emoti-Chair. However, vocalists in the Classical Singer workshop enjoyed feeling the vibrations (M=4.33, SD=0.5) more than the participants in Film-maker workshop (M=3.17, SD=1.47). Further, the vocalists thought that it was not very difficult working with the Emoti-Chair (M=1.77, SD=0.67) while the film-makers believed it to be more difficult (M=3.33, SD=1.5).

Participants in both the workshops agreed that it was easy to perform or compose tactile music. Film artists thought that the task was easier (SM=3.83, SD=1.16) than the singers (M=3.67, SD=0.70). Film artists agreed that there was little difference between auditory and tactile music (M=3.17, SD=1.6) but singers disagreed (M=1.67, SD=1). Both groups agreed that they would need to change how they think about the place of music in their work in order to actually use tactile music; film-makers (M=4.25, SD=0.95), vocalists (M=3.88, SD=1.05).

The film-makers strongly agreed that they wanted to compose or making vibrotactile music for their films (M=4.44, SD=0.89). The vocalists also agreed that they wanted to perform tactile music (M=3.33, S.D=0.70). One of the possible reasons for this difference could be that the film-makers were paid for their participation and may have been compelled to be more positive.

Both the groups wanted to have Emoti-Chairs installed at their future concerts. Moreover, majority of the participants believed that it would take a few months to become proficient at making tactile music.

### 4.2 Thematic analysis of commentary

Examining the comments made by participants, we found that film-makers made an average of 8 comments (SD= 3.265) while the classical singers made an average 13.67 comments (SD= 12.538).

#### 4.2.1 Film-makers' workshop

The deaf film-makers made a total of 48 comments through a sign language interpreter with most comments (14) falling under the improvement category. Figure 3 shows the groupings of all of the comments for each category. Under the 'improvement theme', two of the artists stated that they wanted the low and high frequencies to be labeled so that it would be easier for them to compose vibe-tracks. One female participant suggested translating the entire frequency range of the piano into a range that fit within the tactile sense as it was difficult for her to experience higher frequencies. Another film-maker proposed the use of electrodes on the fingers to convey sound through vibrations in addition to the chair-setup.

The research team is already looking at alternative ways of conveying vibrational information; this is where wearable computing comes into picture. Vibe-vest could be one possible solution similar to [23] but additional research regarding hardware (heat dissipation, power consumption, transmission efficiency) and software is required.

One of the questions raised by participants was, "How would one know what kind of vibration corresponds to which instrument?" This was surprising to the research team as it was explicit in the workshop orientation that the purpose of the vibetrack was to provide a new type of film accompaniment that consists of only vibrations and that the goal was not to translate from the audio to the vibrotactile domain, but to create something entirely new with the vibrotactile domain. Some of the participants eventually understood this point, while other remained stuck on worrying about what certain vibrations would sound like.

The artists thought that the pre-designed vibration blocks were too simple and wanted more complex series of vibrations. Some of the participants added that they wanted a full-body-submerge experience to completely feel the music. Two female participants reported that they were not comfortable with the tickling produced by the vibrations from the Emoti-Chair.

Moreover, one film-maker stated that he expected the vibrations to be smoother and continuous. Regardless, the participants believed that implementing the chair in cinema theatres would be a good option as it would combine the movie watching experience along with the sense of 'feeling' the tactile music over the body.

It was surprising for the researchers to note that a few participants felt disconnected after sitting on the chair because they were unable to concentrate due to the distraction caused by the vibrations. We also observed that participants were challenged to represent sound effects such as thunder or lightning in terms of meaningful vibrations but seemed to better understand the musical representation.

Nonetheless, participants made such statements as "The research is really new and innovative" and "This may lead to a new kind of art form". A majority of the participants showed strong desire to use tactile music and have Emoti-Chairs for their upcoming films. We also noted their desire to have more workshops of longer duration. One of the participants remarked the tactile experience on the Emoti-Chair as technical progress in the deaf culture.

### 4.2.2 Classical Singers' workshop

The responses categorized under the 'vibrational theme' differed in the Classical singer workshop likely because the artists were vocalists and music professors whose approach to and use of music would be quite different from film-makers. Participants from these groups had not necessarily experienced more vibration, but rather, commented more frequently about the physical sensation of the vibration from the Emoti-Chair as their music experience may be a more physical endeavor (e.g., they create the music with their physical body using their voice rather than compile it). As seen in Figure 3, the frequency of comments in the vibrational category in Film-maker and Classical Singer workshops was 7 and 41 respectively.

This difference may also be for other reasons, such as that there were twice as many people in the Classical Singer workshop compared to the one involving film-makers; thereby, increasing the frequency. In addition, the experimental setting of the workshop allowed the classical singers to freely converse with their group members at a relatively faster pace; thereby commenting more often, especially about the sensation of vibration. Film-makers were mostly communicating with the team of hearing researchers through sign language interpreters. The need to work through an interpreter may reduce the comfort levels of participants in communicating with researchers resulting in the fewer comments.

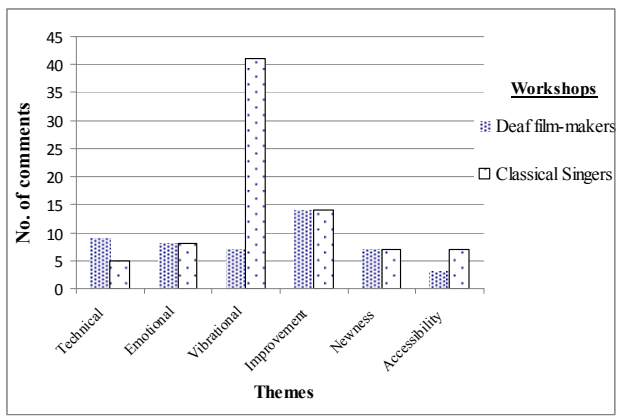

Figure 3: Frequency of thematic comments

In the Classical Singer workshop, most vocalists commented on the vibrational aspects of the chair as they were given the opportunity to sing and sense the vibration generated from their voices. Their experience with the chair was enhanced by the rhythm, rather than the intensity (i.e. crescendo, decrescendo). Participants related the vibrations to some of the properties of music. One participant stated, "The pitch that I sing and hear is the same ones that really translate to the chair." This was also demonstrated by some of the comments in the "newness category", for example "I really loved when you could feel the vibration in two places because of the overtone series; that was by far, my favorite" and "The most interesting side of physical vibration point of view was note being hit". Though the primary purpose of the Emoti-Chair was to translate the musical elements to physical sensation, participants also insinuated having sexual stimulation.

As vocalists, they expressed the idea of adding 'gesture' while singing to visually enhance the experience of the musical elements of the person sitting in the chair. Music, particularly live music, is inherently multisensory and thus it is hypothesized that adding a visual/gesture element would enhance the experience. However, how additional visual/gestural element would change the user experience requires further research. Classical singers also thought that rhythm contributed more to the vibrational experience of music than other music elements such as pitch.

Participants from both workshops suggested the use of electrodes on other body parts such as the hands along with the chair to improve the overall sensation. One participant suggested that this would provide a more "intimate, personal, and complete musical experience." Not surprisingly, participants made several suggestions under 'accessibility theme', which included the use of Emoti-Chair in clinical setting such as speech therapy for the deaf or people with Parkinson's disease who have lost their voices.

As a result of the comments provided in the workshops, it would seem that artists find the notion of composing or using vibrational music intriguing and potentially offering a new opportunity for creative expression and audience experience. They also indicated the desire to learn more about how to use the system and integrate into the work.

## 5 LIMITATIONS

The study provided some insight into the experience of vocalists and movie makers with vibrotactile music, which as seen from results, was largely positive. However, the study also had several important limitations.

First, the number of participants in both the workshops was small. Having more people participate could have given the research team higher certainty about the results. Furthermore, a majority of the participants pointed out that working with others or while being observed was a distraction. Another limitation was that the group involved in the Film-maker workshop comprised only of individuals from the deaf community while the classical singers had no deaf participants. Although, the classical singers could be artificially deafened for the studies, none had any experience being deaf or how a deaf participant may interpret sound through vibration. For future studies, we plan to include more people from both groups.

The video clip used for Toronto participants to create vibrotactile music was from the Harry Potter series. It is possible that some people may have had difficulty in mapping this kind of sound theme to vibration because they either did not work in that genre or did not like it. Using a variety of clips from different genres could have provided greater opportunity for people to express their own creativity.

Last, as the whole concept of composing vibrotactile music was new to participants, longer duration workshops were needed that could have provided participants with additional time to compose and perhaps to provide more in-depth or detailed comments.

## 6 CONCLUSION AND FUTURE WORK

The Emoti-Chair was demonstrated to an audience comprising of professional film-makers, singers and artists. One group composed vibrotactile music, for the first time ever, while the other group experienced their voices through tactile vibrations on The Emoti-Chair. Although the demography of the two workshops was heterogeneous in composition, both groups experienced The Emoti-Chair as a very positive and innovative technology.

The results show that participants were positive about using the chair during the workshops and for future projects. They were particularly interested in including HH/D audiences among their target audiences. Future work could then include allowing artists to explore the compositional space for a deaf or hard of hearing audience and examining the process and audience reaction. We also want to examine how other alternative technology such as visualization could be used to complement the process. Further, we want to construct devices or instruments and structures that are specifically suited to playing/creating vibrotactile music.

## 7 ACKNOWLEDGEMENT

The authors would like to acknowledge the GRAND NCE, Toronto Arts Council and NSERC #18422-2010 for funding the research project. The first author gratefully acknowledges MITACS Globalink for providing summer research internship opportunity in Canada. We thank David Bobier for organizing the Toronto workshop and Rena Sharon for allowing the workshop to take place as part of the Vancouver Song Institute. Finally, we thank all of the participants from the workshops who graciously gave their time and attention.